\documentclass[rsi,twocolumn,nobibnotes]{revtex4-1}
\usepackage{graphicx}
\usepackage{multirow,bigdelim}
\usepackage{bm}
\usepackage{amsmath}
\usepackage{lipsum}
\usepackage[breaklinks=true]{hyperref}

\hypersetup{colorlinks=true,
pdftitle={Stable Faraday laser},
pdfauthor={James Keaveney},
linkcolor=blue,
citecolor=blue,
urlcolor=blue}

\begin{document}

\title{A single-mode external cavity diode laser using an intra-cavity atomic Faraday filter with short-term linewidth $<400$~kHz and long-term stability of $<1$~MHz}
\date{\today}

\author{James Keaveney}
\email{james.keaveney@durham.ac.uk}
\affiliation{Joint Quantum Centre (JQC) Durham-Newcastle, Department of Physics, Durham University, South Road, Durham, DH1 3LE, United Kingdom}
\author{William J. Hamlyn}
\affiliation{Joint Quantum Centre (JQC) Durham-Newcastle, Department of Physics, Durham University, South Road, Durham, DH1 3LE, United Kingdom}
\author{Charles S. Adams}
\affiliation{Joint Quantum Centre (JQC) Durham-Newcastle, Department of Physics, Durham University, South Road, Durham, DH1 3LE, United Kingdom}
\author{Ifan G. Hughes}
\affiliation{Joint Quantum Centre (JQC) Durham-Newcastle, Department of Physics, Durham University, South Road, Durham, DH1 3LE, United Kingdom}

\begin{abstract}
We report on the development of a diode laser system - the `Faraday laser' - using an atomic Faraday filter as the frequency-selective element. In contrast to typical external-cavity diode laser systems which offer tunable output frequency but require additional control systems in order to achieve a stable output frequency, our system only lases at a single frequency, set by the peak transmission frequency of the internal atomic Farady filter. Our system has both short-term and long-term stability of less than 1~MHz, which is less than the natural linewidth of alkali-atomic D-lines, making similar systems suitable for use as a `turn-key' solution for laser cooling experiments.
\end{abstract}

\maketitle


Diode laser systems have for a long time been the workhorse of atomic physics, used in an extremely broad range of applications from laser cooling to thermal vapour experiments~\cite{Wieman1991}. In the usual case of an external cavity diode laser (ECDL), the external cavity is formed using a grating which results ina frequency-selective feedback, ensuring lasing on one dominant mode. Subsequently, their output frequency is stabilised (`locked') using either an external cavity~\cite{Hemmerich1990}, or an atomic frequency reference with techniques such as the dichroic atomic vapour laser lock (DAVLL)~\cite{Corwin1998}, frequency modulation~\cite{Bjorklund1980} and modulation-transfer~\cite{McCarron2008} spectroscopy, polarisation spectroscopy~\cite{pearman02} and electromagnetically-induced-transparency~\cite{Abel2009} to narrow their linewidth and eliminate long-term drift.
Whilst the ease of use and reliability of diode laser systems is continually increasing, the need for locking necessitates the use of external optics adding cost, volume and requiring users to have a high degree of experitise in order to operate such systems successfully. In addition, such systems often still need frequent attention to maintain their stability.

Here we report on the development of a laser system using an atomic vapour Faraday filter as the frequency-selective element - we call this system the `Faraday laser'. 
Atomic Faraday filters have been well studied for over 60 years~\cite{Ohman1956}, and now find applications across many fields (see ref. \cite{Zentile2015a} for a comprehensive literature review).
In comparison with standard ECDL systems, the lasing frequency is limited to the narrow-band transmission window of the combined Faraday filter and cavity system.
Whilst the concept of the Faraday laser was initially developed many years ago~\cite{Wanninger1992,Choi1993} and recently re-evaluated~\cite{Miao2011}, all these works suffer from 
long-term drift of the laser's output frequency. One option of reducing this long-term drift is to use an optical fiber in the external cavity~\cite{Zhu2015,Tao2016}, thus creating an external cavity whose modes are very closely spaced in frequency.
In contrast, we solve the long-term stability problem by using a short cavity, such that only one cavity mode lies under the Faraday filter transmission profile at once, and use standard lock-in techniques to lock the cavity to the peak transmission of the Faraday filter.
Our complete, stabilised laser system is physically no larger than a standard commercial ECDL system, and is a `turn-key' system with essentially no user input required. This would be particularly useful in, for example, laser cooling experiments as a repump laser, or in teaching laboratories where users are non-experts but require a stable laser system.


%
\begin{figure}[b]
\includegraphics[width=0.9\columnwidth,bb=0 0 255 100]{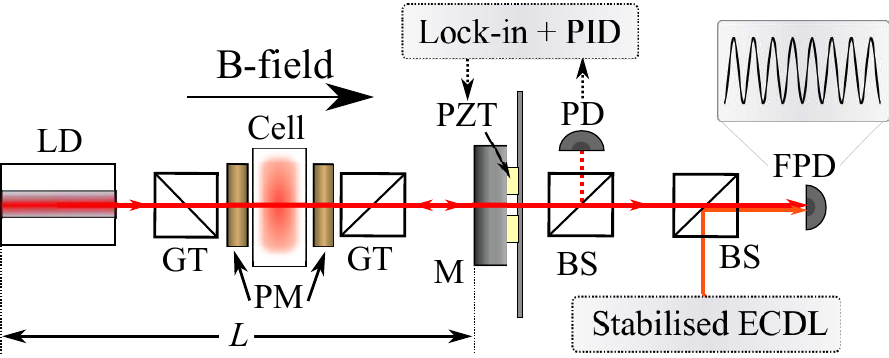}
\caption{Schematic of the Faraday laser design and testing setup. LD - Laser diode; GT - Polarising beam splitter cube; PM - permanent magnet; M - partially reflecting mirror; PZT - piezo transducer; PD - photodiode; BS - beam splitter; FPD - fast photodiode. See main text for discussion of setup.}
\label{fig:setup}
\end{figure}
The basic design of the Faraday laser system is shown in figure~\ref{fig:setup}. The external cavity of the laser, of length $L\sim70$~mm (and hence free-spectral-range of $\sim2.1$~GHz), is formed between the back-facet of the laser diode (LD) and the partially-reflecting ($R=90\%$) mirror (M). The laser diode's output facet is anti-reflection coated, and the beam is collimated with an aspheric lens to form an elliptical beam with $1/e^2$ radii of approximately 0.7 and 2 mm. Inside the cavity, the atomic Faraday filter is formed by two crossed-polarisers (GT) and the anti-reflection-coated vapour cell, with thickness $\ell = 5$~mm. Two Nd permanent magnets (PM) create an axial magnetic field $B$ across the vapour cell. The cavity mirror is mounted on a piezo-electric transducer (PZT), which allows the cavity length to be controlled. We apply a sinusoidal modulation at a frequency of 1 kHz to the PZT, and use a photodiode external to the laser cavity (PD) to monitor the output power. The photodiode signal is the control signal for a PID feedback loop, whereby the piezo voltage is controlled to keep the laser at maximum output power.

The exact operating frequency of the laser is set by the peak of the Faraday filter transmission.
For efficient single-mode lasing, we engineer a single high-transmission filter peak that overlaps with a single cavity mode. 
\begin{figure}[t]
\includegraphics[width=0.9\columnwidth,bb=0 0 310 230]{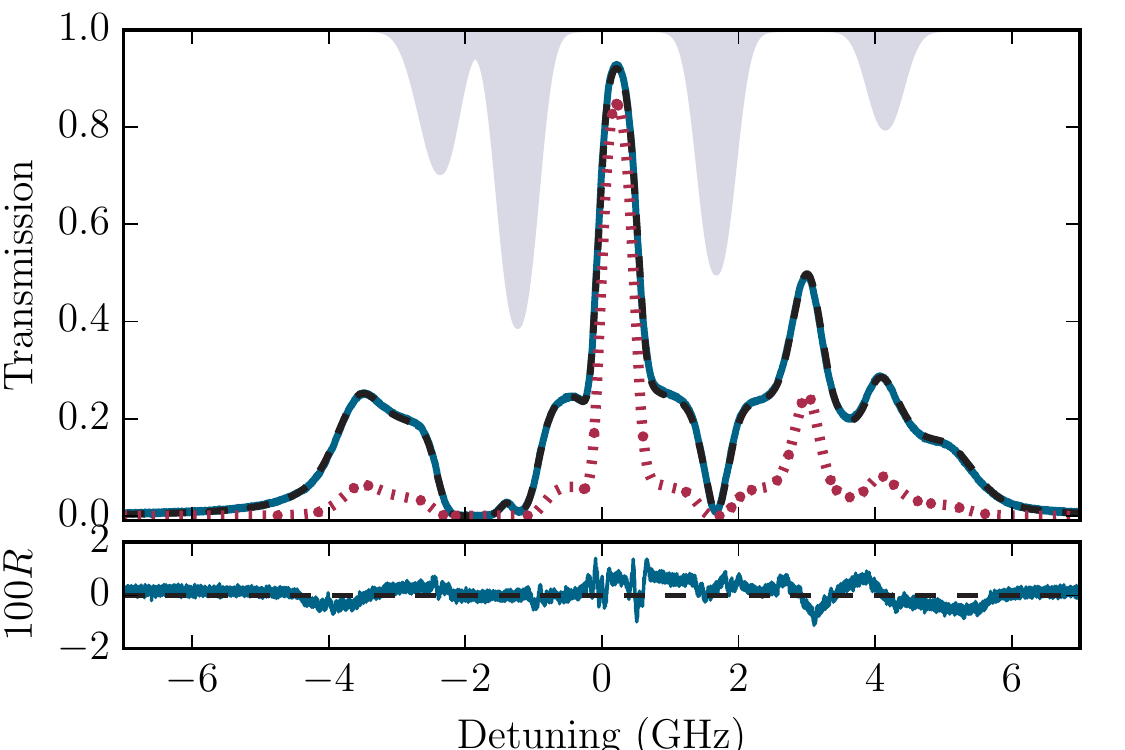}
\caption{Faraday filter transmission as a function of laser frequency detuning. The excellent agreement between experimental data (blue solid) and theory (black dashed) is clear. The bottom panel shows the residuals ($R$), which have been multiplied by 100 for clarity. Fit parameters $B=248.5$~G, $T=90.3^\circ$C. The dashed red line shows the expected double-pass filter transmission. Maximum transmission is 93\% and occurs at $\sim200$~MHz detuning. Zero detuning is defined as the weighted line-centre of the Rb D2 line~\cite{Siddons2008b}. For reference, we also plot the zero-field, room-temperature transmission of a naturally abundant Rb reference cell as a grey outline.}
\label{fig:filter}
\end{figure}
High transmission in Faraday filters occurs where the light is rotated by $\pi/2$, but with low absorptive losses. Modelling this is non-trivial due to the complex hyperfine structure and isotopic abundances in Rb; details can be found in our previous publications~\cite{Siddons2008b,Weller2011a,Keaveney2012,Zentile2015b}. We use the program {\it ElecSus}~\cite{Zentile2015b} to calculate the filter profile in this work.

In practice, once the cell thickness, buffer gas broadening~\cite{Zentile2015a} and atomic abundance ratios have been fixed (by choice of vapour cell), there are only two adjustable parameters; temperature $T$ (which sets the atomic number density), and applied magnetic field $B$. In our system, we use a naturally abundant Rb vapour cell of length 5~mm with no additional buffer gases, and we find high ($>90\%$) maximum transmission at $T=90^{\circ}$C and $B = 250$~G. We choose these conditions as both the field and temperature are easily achievable in the lab, as a proof-of-principle demonstration.
Other operating conditions are available, e.g. the $^{87}$Rb laser cooling or repump frequencies~\cite{Keaveney2016}.

To test the filter performance, we measure the transmission of a second ECDL through the filter as a function of frequency. The Faraday filter spectrum is shown in figure~\ref{fig:filter}. We plot experimental data (blue) and the fitted theoretical spectrum (black dashed line), which show excellent agreement, as indicated by the residuals. The fit parameters are within 1\% of the design parameters of 250~G, 90$^\circ$C. Since the light passes through the cell twice inside the cavity, we also plot the double-pass filter (which is in effect just the filter profile squared) - the main transmission peak remains relatively unchanged (86\% rather than 93\% transmission), whilst the off-band transmission is more dramatically reduced.

With the present design, the output power of the laser is around 0.5~mW with 27~mA drive current (just above lasing threshold), which is enough to seed a second `slave` laser if additional output power is required. The intra-cavity power is $\sim10\times$ the extra-cavity power. Whilst the diode can operate up to 150~mA, the Faraday filter lineshape becomes power-dependent at high intensity, due to optical pumping effects~\cite{Smith2004,Sherlock2009}. When operating the laser in the high-power regime, the long-term (over many hours) laser current drift is the dominant source of instability (current controller: Thorlabs LDC202C), which appears as correlated long-term drifts in the monitored output power and output frequency. For this reason we operate with lower current, where the power-dependence of the Faraday filter is small.

To measure the Faraday laser frequency, we use a commercial ECDL stabilised to the $5S_{1/2} F=3\rightarrow 5P_{3/2} F=4$ transition in $^{85}$Rb using polarisaiton spectroscopy~\cite{pearman02}, and measure the heterodyne beat-note (illlustration shown in fig~\ref{fig:setup}) with the output of the Faraday laser. We measure the beat-note frequency, $f_{\rm BN}$, corresponding to the difference in frequency 
between the two lasers on a fast photodiode (FPD), and monitor this beat-note frequency over time using the FFT function of a fast oscilloscope or, for short-term linewidth measurements, on a spectrum analyser.

\begin{figure}[t]
\includegraphics[width=0.95\columnwidth,trim={0 0.42cm 0 0.45cm},clip, bb=0 0 320 270]{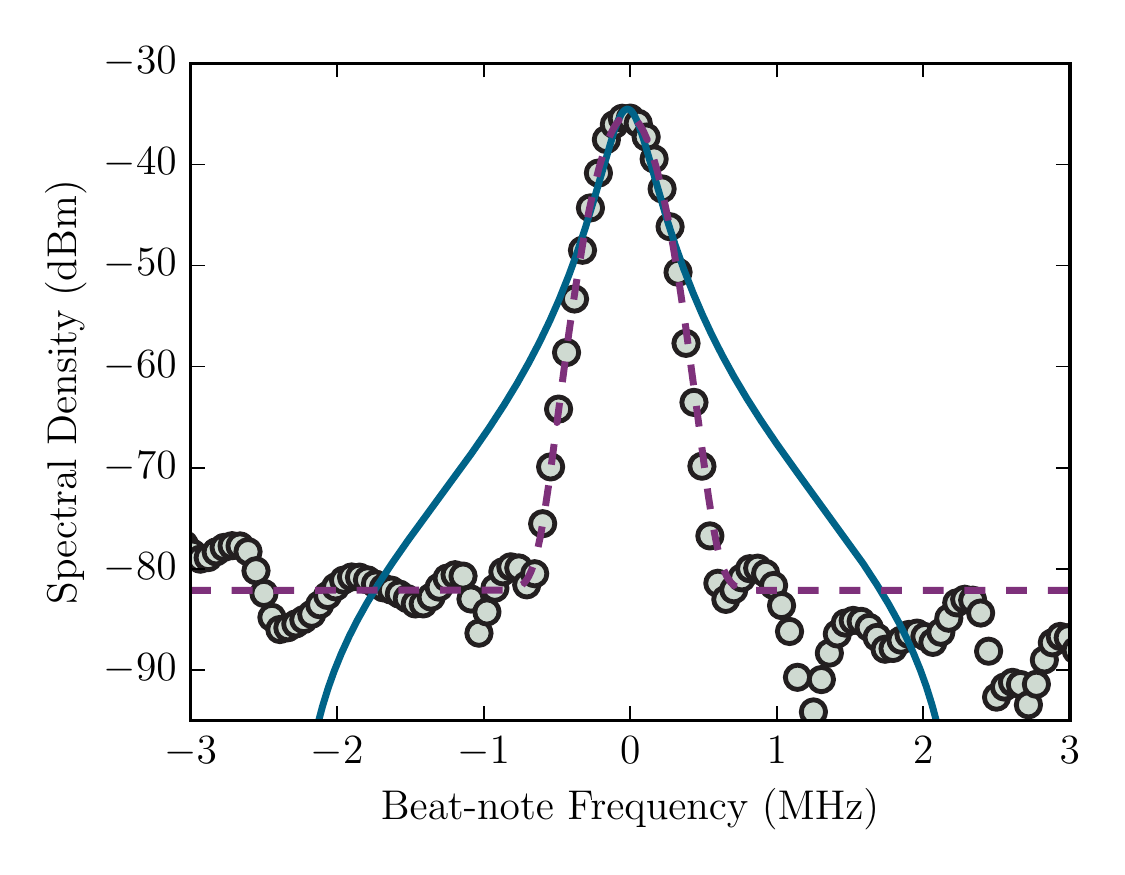}
\caption{Short-time beat-note stability measurement using spectrum analyser. Blue line: Lorentzian fit with FWHM $0.34 \pm 0.01$~MHz, dashed purple line: Gaussian fit with FWHM $0.40 \pm 0.01$~MHz. Resolution bandwidth 3~kHz, video bandwidth 3~kHz, sweep over 25 MHz in 10~ms.
}
\label{fig:linewidth}
\end{figure}

An important measure of laser performance is the short-term linewidth, which we measure on a spectrum analyser. In figure~\ref{fig:linewidth} we show a 10~ms single-shot spectrum of the beat-note frequency (relative to the central operating frequency) and fit the data to both Gaussian and Lorentzian line shapes.
The data fit well to a Gaussian (dashed purple line) with full-width-half-maximum (FWHM) of $0.40\pm0.01$~MHz. That the data do not fit well to a Lorentzian (blue line)
indicates that technical noise (pink-noise, $1/f$) is likely to be the dominant source of line-broadening~\cite{Turner2002}. The exact source of this noise is beyond the scope of the current investigation. We note that this is the combined linewidth of both laser systems - the measured linewidth is the convolution of the linewidth of each laser.
The reported linewidth is therefore an upper bound on the short-term linewidth of the Faraday laser.
\begin{figure}[t]
\includegraphics[width=0.95\columnwidth,trim={0 0.42cm 0 0.45cm},clip,bb=0 0 310 270]{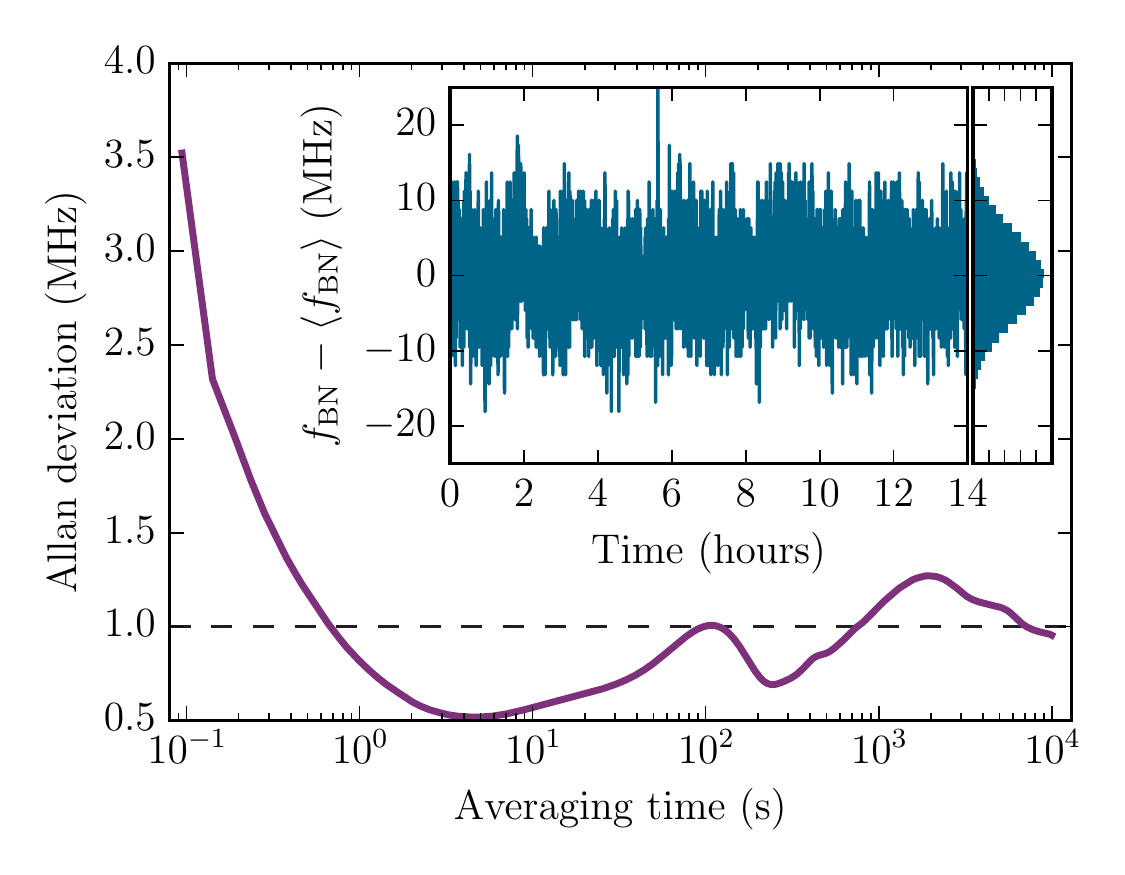}
\caption{Long-term beat-note stability measurement as measured by the Allan deviation. Raw data are shown in the inset; to indicate the scatter in data points, we also plot a histogram of the data values.}
\label{fig:stability}
\end{figure}

With only passive frequency stability controls, the laser would drift over time as the cavity length drifts. To combat this, we employ active stabilisation by modulating the cavity length via the PZT and detecting the laser output on a photodiode. We form an error signal via a lock-in amplifier which is passed to a PID controller, locking the laser output frequency to the peak of the Faraday filter transmission curve.
We monitor long-term frequency stability using the fast Fourier transform (FFT) function of a fast oscilloscope (2.5 GHz bandwidth). Each oscilloscope trace is sampled at 20 GS/s with a 40~kS record length, at a repetition rate of approximately 20 Hz over many hours, so it is impractical to save all the data. 
%
%
Instead, we record the frequency of the peak FFT intensity. We then analyse this data set using the standard approach of an overlapping Allan deviation~\cite{Allan1966,NIST2008}. In figure~\ref{fig:stability} we plot the Allan deviation as a function of the averaging time, and in the inset we show an example of the raw frequency data (average beat-note frequency removed for clarity). The full data set is over 1 million points (only a small fraction is plotted). 

With a single detector, we are limited to timescales longer than half the sampling rate of the oscilloscope ($\tau \gtrsim 100$~ms).
We could improve the short-time resolution by using another detector~\cite{Kunze1996}, but this is beyond the scope of this investigation. The Allan deviation data show that the frequency stability of the laser is less than 1 MHz (dashed horizontal line) for nearly all time scales greater than 1 second, which is less than the linewidth of all alkali-atom D-lines, making similar systems (operating at slightly different wavelengths) suitable for applications in laser cooling.

In conclusion, we have developed a proof-of-principle stable laser system with intrinsic atomic feedback, reducing the need for external optics and expertise that is usually required to frequency-stabilise an ECDL operating near an alkali-atom D-line transition.
With the current design, we achieve both short-term linewidth and long-term frequency stability of better than 1 MHz.
Similar high-transmission Faraday filters can be engineered for alternate operating wavelengths for the other alkali-metal atoms, so in principle this type of design is generally applicable to all alkali-atoms.

The authors would like to thank D. J. Whiting, M. A. Zentile and E. M. Bridge for fruiful discussions, and acknowledge funding from EPSRC (Grant No. EP/L023024/1), the Ogden Trust and Durham University. The data presented in this paper are available from \footnote{\url{http://dx.doi.org/10.15128/r1dn39x1523}}.
 
\bibliography{library,library2}

\end{document}